\documentclass[11pt]{article}
\usepackage{moriond,epsfig}

\def\Journal#1#2#3#4{{#1} {\bf #2}, #3 (#4)}

\def\PRL{\em Phys. Rev. Lett.}
\def\PRD{{\em Phys. Rev.} D}

\def\dpipi{\ensuremath{D^0\to \pi^+\pi^-}}
\def\dKK{\ensuremath{D^0\to K^+K^-}}
\def\dAcp{\ensuremath{\Delta A_{\rm{CP}}}}

\def\phis{\ensuremath{\phi_s}}
\def\DGs{\ensuremath{\Delta\Gamma_s}}
\def\Bs{\ensuremath{B^0_s}}
\def\B0{\ensuremath{B^0}}
\def\bBs{\ensuremath{\bar{B}^0_s}}
\def\BsMix{\Bs--\bBs}
\def\jpsiphi{\ensuremath{\Bs \to J\!/\!\psi \varphi}}
\def\jpsi{\ensuremath{J\!/\!\psi}}
\def\bsmumu{\ensuremath{\Bs \to \mu^+ \mu^-}}
\def\bdmumu{\ensuremath{\B0 \to \mu^+ \mu^-}}
\begin{document}
\vspace*{1cm}
\title{RECENT HEAVY FLAVOR RESULTS FROM THE TEVATRON}

\author{M. DORIGO on behalf of the CDF and D0 Collaborations.}

\address{University of Trieste and INFN Trieste}

\maketitle\abstracts{
The CDF and D0 experiments at the Tevatron $p\bar{p}$ collider have pioneered and established 
the role of flavor physics in hadron collisions. 
A broad program is now at its full maturity.
We report on three new results sensitive to physics
beyond the standard model, obtained using the whole CDF dataset:
a measurement of the difference of CP asymmetries in $K^+K^-$ and $\pi^+\pi^-$ decays of $D^0$ mesons, 
 new bounds on the \Bs\ mixing phase and on the decay width difference of \Bs\ mass-eigenstates, and
an update of the summer 2011 search for $\B0_{(s)}$ mesons decaying into pairs of muons.	
Finally,  the D0 confirmation of the observation 
of a new hadron, the $\chi_b(3P)$ state, is briefly mentioned.}

\section{Measurement of CP violation in charm in the final CDF dataset}
Violation of the CP symmetry in tree-dominated decays \dpipi\ and  \dKK\  
is a sensitive probe of physics beyond the standard model (SM).
Both the $D^0$--$\bar{D}^0$ mixing amplitude and the SM-suppressed penguin amplitude
can be greatly enhanced by new dynamics, 
which can also increase the size of the CP violation over that expected from
to the Cabibbo-Kobayashi-Maskawa (CKM) hierarchy.
Last year, using 5.9 fb$^{-1}$ of data, CDF produced the world's most precise measurements of 
the CP asymmetries $A_{\rm{CP}}(KK)=(-0.24\pm0.22\pm0.09)\%$ and  $A_{\rm{CP}}(\pi\pi)=(0.22\pm0.24\pm0.11)\%$.\cite{Acp_cdf}
In spite of the hadronic uncertainties, there is broad consensus that direct CP asymmetries of \dKK\ and of \dpipi\ should be of opposite sign. 
Therefore, a measurement of the difference between asymmetries of those decays
 is maximally sensitive to detect CP violation. 
Indeed, the LHCb collaboration reported recently the first evidence of CP violation in charm 
measuring $\dAcp =A_{\rm{CP}}(KK)-A_{\rm{CP}}(\pi\pi)= (-0.82 \pm 0.21 \pm 0.11)\%$.\cite{dAcp_lhcb}
An independent measurement is crucial to establish the effect, 
 and the 10 fb$^{-1}$ sample of hadronic $D$ decays collected by CDF is the only one currently available 
to attain sufficient precision. 

\begin{figure}
\begin{center}
\includegraphics[width=0.47\textwidth]{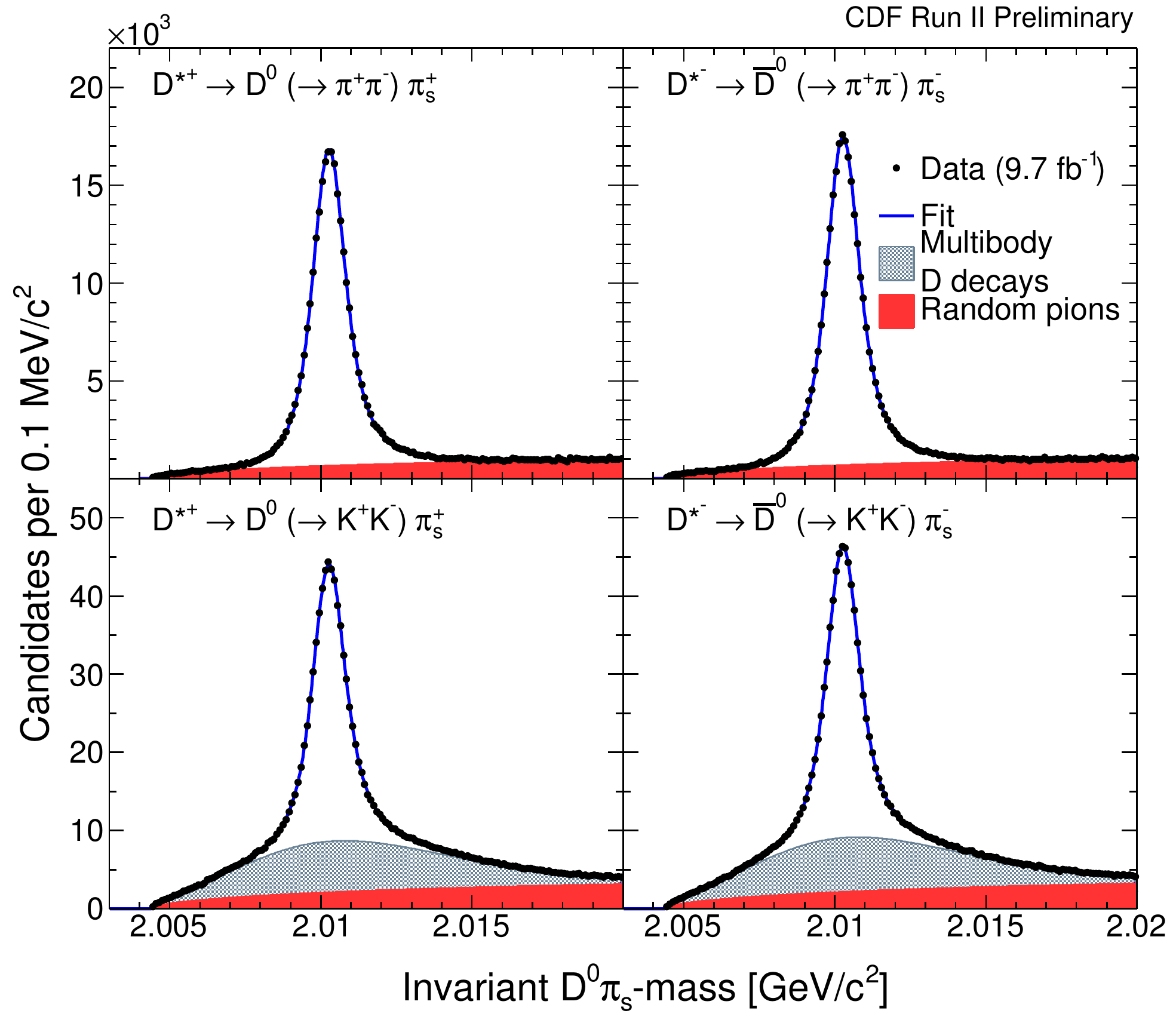} \hspace{10 mm}
\includegraphics[width=0.45\textwidth]{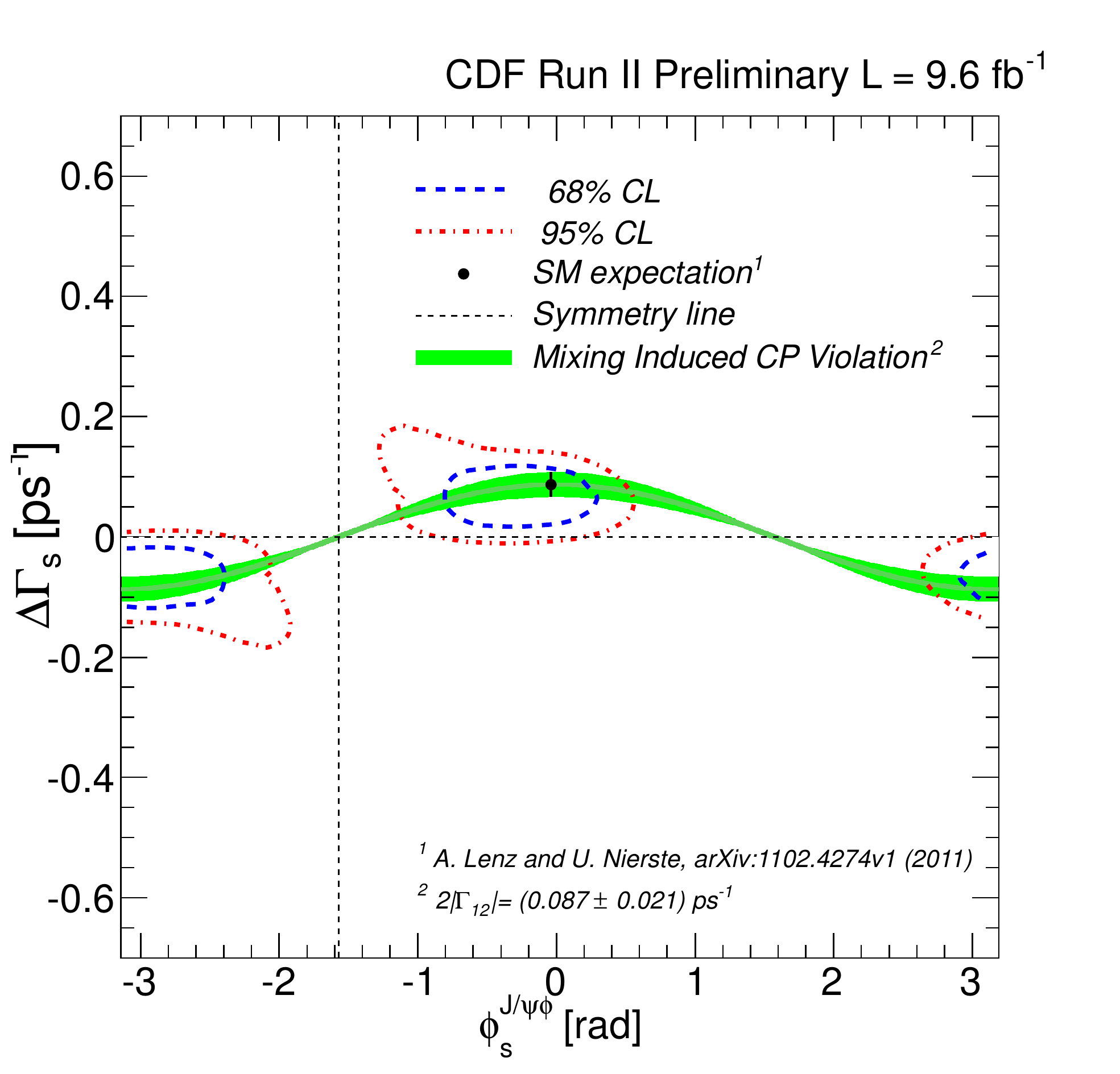}
\end{center}
\caption{Left: $D^0\pi$-mass
distributions of positive and negative $D^\star$ decays with fit projections overlaid.\hspace{10mm} Right: 68\%  and 95\% confidence regions in the plane 
(\phis,\DGs) from profile-likelihood of CDF data.}
\label{fig:charm_asymmetries}
\end{figure}

The analysis follows closely the measurement of individual asymmetries.\cite{dAcp_cdf}
The flavor of the $D^0$ meson is tagged from the charge of the soft pion in the 
strong $D^{\star +} \to D^0 \pi^+$ decay. Since $D^{\star +}$ and $D^{\star -}$ mesons are produced 
in equal number in strong $p\bar{p}$ interactions, any asymmetry between 
the number of $D^0$ and $\bar{D}^0$ decays is due to either CP violation or instrumental effects.
The latter can be induced only by the difference in reconstruction
efficiency between positive and negative soft pions. Provided that the relevant kinematic distributions 
are equalized in the two decay channels, the instrumental asymmetry cancels 
to an excellent level of accuracy in the difference between the observed asymmetries between signal yields.
Such cancellation allows one to increase the sensitivity on \dAcp\ by loosening some selection criteria with respect to 
the measurement of individual asymmetries and double the signal yields. 
The trigger is fired by two charged particles with transverse momenta greater than 2 GeV/$c$.
The excellent CDF momentum resolution yields precise mass resolution ($\sim$8 MeV/$c^2$ for $D$ mesons)  
which provides good signal-to-background.
The typical resolution (50 $\mu$m) on the impact parameter (IP) of the tracks
is effective to trigger on good kaon or pion candidates,  
that are required to have IP$>$100 $\mu$m.
The offline selection adds some basic additional requirements 
on track and vertex quality.  
The numbers of $D^0$ and $\bar{D}^0$ decays are determined with a simultaneous fit to the $D^0\pi$-mass
distribution of positive and negative $D^\star$ decays (see Fig.~\ref{fig:charm_asymmetries}, left). 
 About 1.21 million \dKK\ 
decays and 550 thousand \dpipi\ decays are reconstructed, yielding the following 
observed asymmetries between signal yields, 
$A_{\rm{raw}}(KK)=(-2.33\pm0.14)\%$ and $A_{\rm{raw}}(\pi\pi)=(-1.71\pm0.15)\%$.
Residual systematic uncertainties 
total 0.10\% and are driven by differences 
between distributions associated with charm and anticharm decays. 
The final result is $\dAcp=(-0.62\pm0.21\pm0.10)\%$, 
which is 2.7$\sigma$ different from zero.
This provides strong indication of CP violation in CDF charm data, 
supporting the LHCb earlier evidence with same resolution. 
 The combination of CDF, LHCb and $B$-factories measurements  
deviates by approximately 3.8$\sigma$ from the no CP violation point.

\section{Measurement of the \jpsiphi\ time-evolution in the final CDF dataset}
The \BsMix\ mixing is a promising process where to search for new physics (NP), 
given the D0 3.9$\sigma$ anomaly in dimuon charge asymmetry.\cite{Asl_d0} 
 If the anomaly is due to new dynamics in the \Bs\ sector, 
the phase difference between the \BsMix\ mixing amplitude
and the amplitude of  \Bs\ and \bBs\ decays into common 
final states, \phis, would be significantly  
altered with respect to its nearly vanishing value expected in the SM. 
A non-CKM enhancement of \phis\ can also decrease 
the decay width difference between the heavy 
and light mass-eigenstate of the \Bs\ meson, \DGs.
The analysis of the time evolution of \jpsiphi\ decays 
is the most effective experimental probe of such a CP-violating phase. 
Since the decay is dominated by a single real amplitude, 
the phase difference equals the mixing phase to a good approximation. 
Early Tevatron measurements have shown a mild discrepancy 
of about 2$\sigma$ with the SM expectation.\cite{phis_cdf_d0_comb} Latest updates by CDF and D0 
are in better agreement with the SM, as well as first measurements provided by LHCb.\cite{phis_cdf_5fb,phis_d0_8fb,phis_lhcb_prl} 

Here we report the new CDF update using the final dataset of 10 fb$^{-1}$ which comprises
about 11000 $\jpsiphi$ decays collected by a low-momenta dimuon trigger.\cite{phis_cdf_new} 
The decays are fully reconstructed through four tracks that fit to a common displaced vertex, 
two matched to muon pairs consistent with a \jpsi\ decay, and two consistent with a $\phi \to K^+K^-$ decay.
  A joint fit that exploits the candidate-specific information given by the 
  $B$ mass, decay time and production flavor, along with the 
 decay angles of kaons and muons, is used to determine both \phis\ and \DGs. 
 The analysis closely follows the previous measurement obtained on a subset of the data.\cite{phis_cdf_5fb}
 The only major difference is the use of an updated calibration of  the 
 tagging algorithm that uses information from the decay of the ``opposite side" bottom hadron
 in the event to determine the flavor of the \Bs\ at its production, with tagging power $(1.39 \pm 0.01)\%$. 
 The information from the tagger that exploits charge-flavor correlations of
 the neighboring kaon to the \Bs\ is instead restricted to only half of the sample, in which has tagging power $(3.2 \pm 1.4)\%$. 
 This degrades the statistical resolution on \phis\ by no more than 15\%.
 A decay-resolution of $\sim$90 fs allows resolving 
 the fast \Bs\ oscillations to increase sensitivity on the mixing phase. 
The 68\% and 95\% confidence regions in the plane $(\phis,\DGs)$
 obtained from the profile-likelihood of the CDF data are reported in Fig.~\ref{fig:charm_asymmetries} (right).   
The confidence interval for the mixing phase is
 $\phis \in [-0.60,0.12]$ rad at 68\% C.L.,
 in agreement with the CKM value and recent D0 and LHCb determinations.\cite{phis_d0_8fb,phis_lhcb}
 This is the final CDF measurement on the \Bs\ mixing phase, and provides a factor 35\% improvement 
 in resolution with respect to the latest determination.   
CDF also reports 
$\DGs=(0.068 \pm 0.026 \pm 0.007)$ ps$^{-1}$ under the hypothesis of a SM value for \phis,
 along with the measurement of the \Bs\ lifetime, $\tau_s = (1.528 \pm 0.019 \pm 0.009)$ ps, 
 in agreement with other experiments' results.\cite{phis_d0_8fb,phis_lhcb}
 
\section{Final search for dimuon decays of $B$ mesons at CDF}
The $\B0_{(s)} \to \mu^+\mu^-$ decays involve flavor changing neutral currents and 
the observation of an abnormal decay rate can provide excellent evidence of NP 
since in the SM they can occur only through high-order loop diagrams. 
 Enhancements to the SM expectation of their branching ratios (BR) can indeed occur in a variety of different NP models.  
Last summer CDF reported  
an intriguing $\sim$2.5$\sigma$ fluctuation over background in 7 fb$^{-1}$ of data. 
Even if compatible with the SM and other experiments' results, 
it allowed the first two sided bound on the \bsmumu\ rate.\cite{bmm_cdf_7fb}

\begin{figure}
\begin{center}
\includegraphics[width=0.535\textwidth]{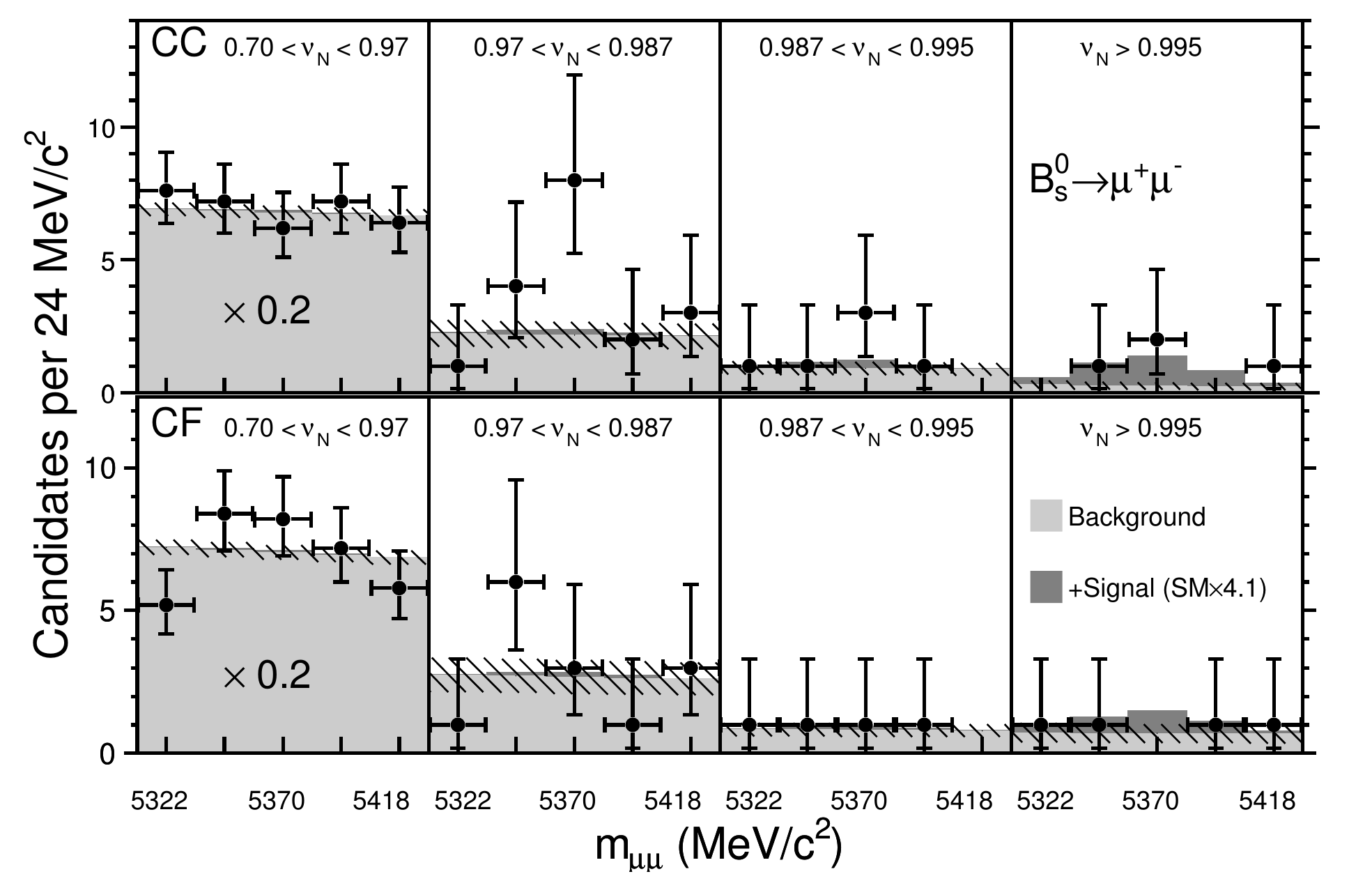}\put(-170,155){{\small CDF Run II Preliminary}} 
\hspace{10 mm}
\includegraphics[width=0.365\textwidth]{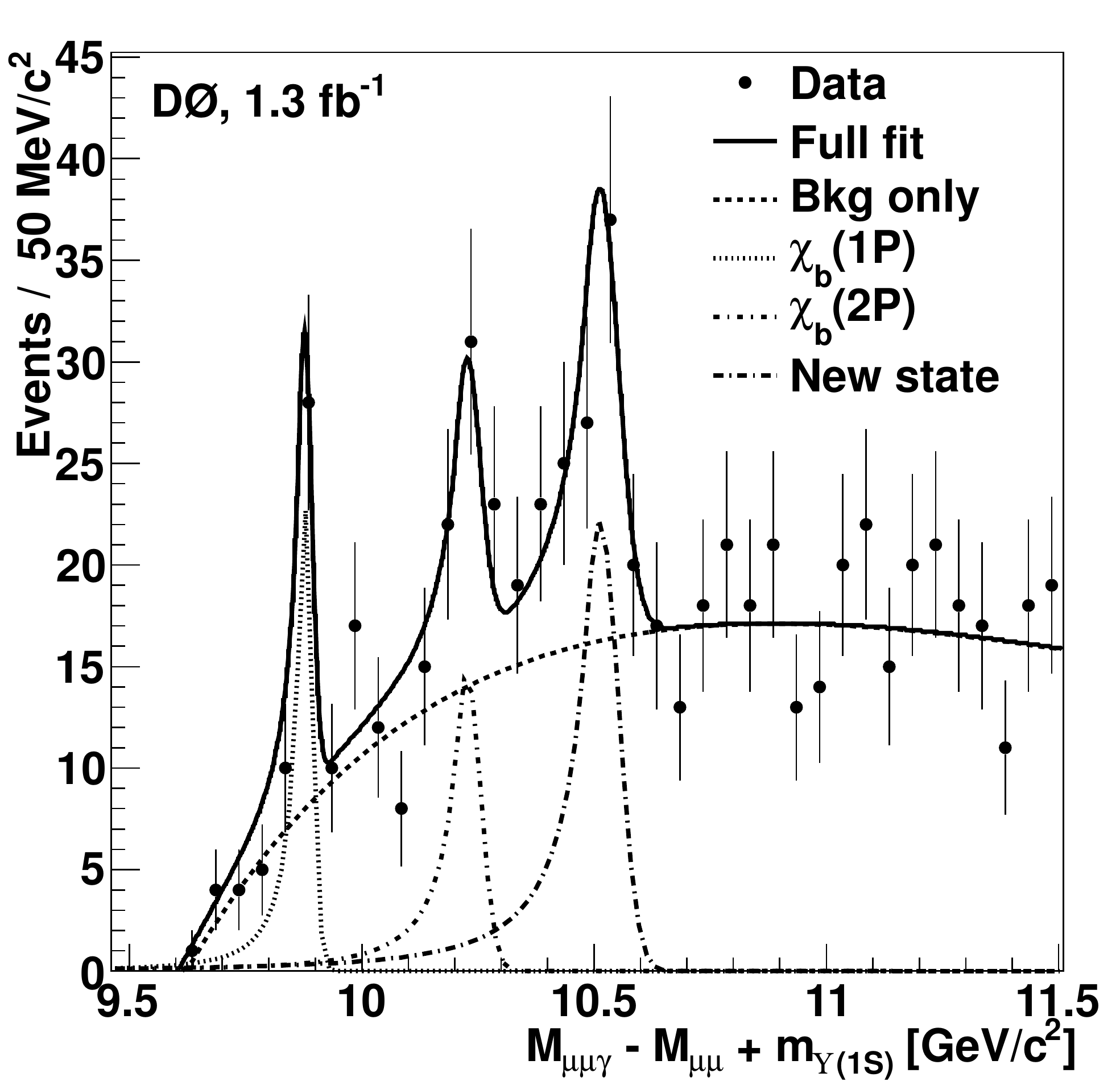}
\end{center}
\caption{Left: comparison of the CDF \bsmumu\ data (points ) and expected background (solid grey) for CC and CF muons. 
Right: mass distribution of $M_{\mu\mu\gamma} - M_{\mu\mu} + M_{\Upsilon(1S)}$ of D0 data with fit overlaid.}
\label{fig:bs_mumu}
\end{figure}

Here we report the CDF update of the analysis with the final 10 fb$^{-1}$ dataset.\cite{bmm_cdf_new} 
The analysis methods are not changed from the previous iteration 
to ensure the unbiased processing of new data. 
The events are collected using a set of dimuon triggers and are divided in two categories:
``CC" events have both muon candidates in the central region of the detector, 
 while ``CF" events have one central muon and another muon in the forward region.
The signal candidates are fully reconstructed with a
secondary vertex due to the long \Bs\ lifetime ($\sim$450 $\mu$m). 
They also feature a primary-to-secondary
vertex vector aligned with the \Bs\ candidate momentum and a very isolated \Bs\ candidate.
There are two sources of background: 
combinatorial background and peaking
background. 
The former tends to be partially reconstructed 
and shorter-lived than signal. It is
estimated by extrapolating the number of events in the sideband regions 
of the $B$ mass distribution to the signal window using a linear fit.
The peaking background are due to decays of \Bs\ and \B0\ mesons
to pions and kaons that are misreconstructed as muons, and it is ten times greater
in the \B0\ search with respect to the \Bs\ analysis. It 
is carefully taken into account with simulation (mass shape) and with 
 $D^0 \to K \pi$ decays from data (misidentification probability).
A neural network (NN) classifier is optimized to reject
the background using 14 event variables.
The background estimates are checked to be consistent in many control samples.
Finally, when the background is
well understood, the number of observed events is compared to the number expected. 
The data are found to be consistent with the background expectations for the \bdmumu\ decay and yield an observed 
limit of $\rm{BR} < 4.6 \times 10^{-9}$ at 95\% C.L (with expected limit $4.2 \times 10^{-9}$). 
In the case of \bsmumu\ the summer 2011 excess is not reinforced, even though it is still present
as shown in the most sensitive (top-right) bin of the NN in Fig.~\ref{fig:bs_mumu}. 
The resulting bounds at 95\%  C.L. are 
$0.8 \times 10^{-9} < \rm{BR} < 3.4 \times 10^{-8}$, which is still compatible both with the SM expectation 
and the latest limits from LHC experiments.\cite{bmm_exp}
An upper limit at 95\% C.L. of $\rm{BR} < 3.1 \times 10^{-8}$ (expected $1.3 \times 10^{-8}$) is also derived.

\section{A new state decaying into $\Upsilon(1S)+\gamma$}
Using data corresponding to an integrated luminosity of 1.3 fb$^{-1}$, 
the D0 collaboration observes a narrow mass state decaying into $\Upsilon(1S) + \gamma$, 
where the $\Upsilon(1S)$ meson is detected by its decay into a muons pair, 
and the photon through its conversion into an $e^+e^-$ pair.\cite{chib3P_d0} 
The fit to the mass spectrum in Fig.~\ref{fig:bs_mumu} (right) shows three structures 
above a smooth, threshold-like background. 
The one at the highest mass has a statistical significance of 5.6$\sigma$. It
 is interpreted as the state $\chi_b(3P)$ and is centered at $10.551 \pm 0.014 \pm 0.017$ GeV/$c^2$, 
 consistent with the recent ATLAS observation.\cite{chib3P_atlas}

\section{Conclusions}
We reported the final CDF results on three flagship measurements in the indirect search for NP at Tevatron, 
and the confirmation of a new hadron ($\chi_b(3P)$) by the D0 collaboration.  
The CP asymmetries of $D^0$ meson decays reported by CDF confirm the first evidence
of CP violation in charm reported by LHCb; in the \Bs\ sector,  
tensions with SM predictions are now softened by latest updates of \phis\ and \DGs\ bounds, 
making the D0 $A_{sl}$ anomaly even harder to depuzzle.
The final CDF update on the \bsmumu\ rate is concluding 
 a decade-long program of Tevatron searches 
that improved the experimental bounds on the rate down to the $10^{-8}$ range, nearing now the 
sensitivity to observe a SM signal. 
Analyses of the unique ($p\bar{p}$ charge-symmetric),  rich ({\it e.g.}, millions of charm decays),
and well-understood (10-years expertise) data sample acquired by CDF and D0 are still in progress and may reserve
interesting results in the near future. 

\section*{References}
\small{

}

\end{document}